
\magnification=1200
\line{\hskip 4 in UTHEP-233 \hfill}
\line{\hskip 4 in April, 1992 \hfill}
\bigskip
\bigskip
\centerline{ \bf On the Renormalization of the S Parameter}
\bigskip
\bigskip
\centerline{ Sinya Aoki$^\diamond$}
\bigskip
\bigskip
\centerline{  Institute of Physics, University of Tsukuba,
Tsukuba, Ibaraki 305, Japan}
\smallskip
\advance
\baselineskip by 0.5
\baselineskip
\bigskip
\vfill
\centerline{ ABSTRACT}
\bigskip
We calculate the S parameter of the standard
model at one loop of fermions, using
three different regularizations (dimensional, Pauli-Villars and
lattice) and find an extra contribution to the S parameter besides
the standard one for each case. This shows
that the extra contribution recently reported for the lattice regularization
is {\it not} necessarily
tied to the non-decoupling effect of fermion doublers.
We argue that the
extra contribution should be subtracted in the renormalizable perturbative
expansion.

\vskip 1.5 in
\vfill
\leftline{$\diamond$ \quad  email: aoki@athena.ph.tsukuba.ac.jp}
\eject
\def\pl#1#2#3{{\it Phys. Lett.} {\bf B#1}(#2)#3}
\def\np#1#2#3{{\it Nucl. Phys.} {\bf B#1}(#2)#3}

\def\pr#1#2#3{{\it Phys. Rev.} {\bf D#1}(#2)#3}
\def\prl#1#2#3{{\it Phys. Rev. Lett.} {\bf #1}(#2)#3}
\def\bar#1{\overline#1}

A successful construction of chiral gauge theories on the lattice
requires a dynamical decoupling of
fermion doublers in the continuum limit.
It now seems that in the
Wilson-Yukawa formulation$^1$ fermion doublers are indeed decoupled in the
region of strong Wilson-Yukawa coupling $^2$.
However, in a recent paper$^3$ by Dugan and Randall, it is found that
the S parameter$^4$ calculated in the Wilson-Yukawa formulation has an extra
contribution
added to the standard continuum value$^{3,4}$.
It is claimed that this extra contribution arises from
fermion doublers, so that a decoupling of doublers does not hold
in the Wilson-Yukawa formulation.
In this paper, however, we point out that an extra contribution also
arises in continuum regularization schemes.
Therefore, an extra contribution is {\it not} necessarily tied to
non-decoupling effects of fermion doublers.

\vskip 0.5cm

The S parameter$^{3,4}$ is defined by
$$
S = - 16\pi
{\partial \Pi_{3Y} (p^2) \over \partial p^2} \vert_{p^2=0} \eqno(1)
$$
where $\Pi_{3Y} (p^2)$ is the part of the ($T^3_L$ current) - ($Y$ current)
correlation function proportional to $\delta_{\mu\nu}$. For a quick review, see
ref. 3. We consider a one-loop contribution of one fermion doublet to this S
parameter
in the standard model. At the fermion 1-loop level, for simplicity,
we take the doublet Higgs field $\phi$ as
$\phi (x) = g(x) \displaystyle {0\choose v} $
where $v$ is a vacuum expectation value of $\phi$ and
$g^\dagger g = 1$ .
The Lagrangian for one fermion doublet in the standard model is given by
$$
L_F = L_0 + L_{reg}  \eqno(2)
$$
where
$$
L_0  =  \bar\psi(x) \sum_{\mu =1}^4\gamma^\mu\ D_\mu \psi(x)
+ yv\  \bar\psi(x)  g^R(x)\cdot g^L(x) \psi(x) ,
\eqno(3)
$$
$$\eqalign{
D_\mu &  = D^L_\mu P_L + D^R_\mu P_R, \quad
D^L_\mu = \partial_\mu - {i\over 2}(g_2 W_\mu^a \tau^a + g_1 Y_L B_\mu), \quad
D^R_\mu = \partial_\mu - {i\over 2}(g_1 Y_R B_\mu) \cr
& \qquad g^L(x) = g^\dagger (x) P_L + P_R, \qquad
g^R(x) = P_L + g(x) P_R  \cr}  \eqno(4)
$$
(some modifications to this short-handed notation should be understood in the
case of the lattice regularization). Here the Yukawa coupling
$y$ is taken to be equal for the both up and down components of the doublet,
$W_\mu^a$ is the SU(2) gauge field, $B_\mu$ is the
hyper-charge U(1) gauge field, $P_{L,R}$ is the chiral projection, and
$Y_{L,R}$ which satisfies $Y_R - Y_L = \tau^3$ is the hyper-charge of
the fermion.

The {\it regulator} part of the Lagrangian is given by
$$\eqalignno{
L_{reg}^{(dim)} & =  \bar\psi \{ g^R  \sum_{\mu = 5}^D\gamma^\mu\ D_\mu
g^L \} \psi  & (5-a) \cr
L_{reg}^{(PV)} & = \sum_i \bar\psi_i \{ \gamma\cdot D +
(yv+M_i) g^R\cdot g^L \} \psi_i  & (5-b) \cr
L_{reg}^{(lat)} & = -{ar}\sum_\mu \bar\psi \{ g^R\ D^R_\mu D^R_\mu
\ g^L\} \psi .  & (5-c) \cr }
$$
Here (5-a) is
for the dimensional regularization$^5$ in $D$ dimensions with the
'tHooft-Veltman
definition of $\gamma_5$ which satisfies $\{\gamma_5,\gamma_\nu\} = 0$ for
$\nu = 1\sim 4$ and $[\gamma_5,\gamma_\nu] = 0$ for $\nu = 5 \sim D$,
(5-b) is
for the Pauli-Villars regularization, where Pauli-Villars fields $\psi_i$
are all ghost fields and have integer weight $C_i$ and mass $M_i$,
and (5-c) is
for the lattice regularization with the Wilson Yukawa term$^{1,6}$
with $a$ the lattice spacing and $r$
the Wilson parameter which is dimensionless and non-zero.

We have to put the Nambu-Goldstone part of Higgs fields, $g$,
in the above Lagrangian in order to
keep gauge invariance for the regularized theory, since the regulator part
connects the left-handed fermion to the right-handed fermion like a
fermion mass term. In this gauge invariant scheme
gauge anomalies appear as Wess-Zumino terms$^7$.
The gauge invariance allows us to take the unitary gauge ( $g=1$ )
for our calculation.
However, because of the non-linearity of $g$,
this scheme is not manifestly perturbatively renormalizable even
for anomaly free theories.

We denote the fermion 1-loop contribution to the S parameter as
$$
S^{1-loop} = S_0^{1-loop} + \delta S^{1-loop} \eqno(6)
$$
where
$
S_0^{1-loop} = \displaystyle {1\over 6\pi}
$
is the value for one doublet
obtained by the momentum cut-off or the dimensional
regularization with anti-commuting $\gamma_5$.
It should be noted that these regularizations are problematical:
The momentum cut-off is difficult to formulate consistently at higher loops and
the anti-commuting $\gamma_5$ fails to
produce {\it global} anomalies as well as gauge anomalies.
The extra contribution $\delta S^{1-loop}$ is non-zero
for the consistent regularizations given in (5).

In the case of the dimensional regularization$^8$,
the correlation function
for $W_\mu^3$ and $B_\nu$ ( with the unitary gauge $g(x)=1$ )
in the momentum space takes the form;
$$\eqalign{
< W_\mu^3\cdot B_\nu > (p) & = \delta_{\mu\nu}
{g_1g_2\over 16\pi^2}{tr (\tau^3 Y_R)\over 2} \times \cr
 & \left[ m^2 \{ {1\over \epsilon}+\ln (4\pi) - \ln ({m\over \mu_0})^2
 -\int_0^1 dt \ln [1+{p^2\over m^2}t(1-t)] \} + m^2 + {p^2\over 6} \right] \cr
& = \delta_{\mu\nu}
{g_1g_2\over 16\pi^2}
\left[ m^2 \{ {1\over \epsilon}+\ln (4\pi) - \ln ({m\over \mu_0})^2
-{p^2\over 6 m^2} \} + m^2 + {p^2\over 6} \right] + O(p^4) \cr } \eqno(7)
$$
where $D = 4 -2\epsilon$, $m = y v$ and $\mu_0$ is some mass scale. From
the first terms which are propotional to $m^2$ we obtain  $S_0^{1-loop}
= \displaystyle {1\over 6\pi}$ and from the last term we find$^8$
$$
\delta S^{1-loop} = -{1\over 6\pi} .  \eqno(8)
$$
For the Pauli-Villars regularization$^8$ a similar calculation yields
$$
\delta S^{1-loop} = -{1\over 6\pi}\sum_i C_i = -{1\over 6\pi}  \eqno(9)
$$
which happens to be equal to the value in the case of the dimensional
regularization because $\sum_i C_i = 1$.
The value of $\delta S^{1-loop}$ for the lattice regularization was
calculated in ref. 3:
$$
\delta S^{1-loop} = F(r) \eqno(10)
$$
where $F(r)$ is a function of $r$ which satisfies
$$
F(r) = \cases { \displaystyle 0 , & for $r\rightarrow 0$ \cr
\displaystyle 1/ 8\pi , & for
$r \rightarrow \infty$  \cr}. \eqno(11)
$$
Here we take the $r\rightarrow 0$ limit {\it after} the $a\rightarrow 0$ limit.
See ref. 3 for an explicit form of $F(r)$ .

Now it becomes clear that the existence of a non-zero $\delta S$ does not
mean non-decoupling of fermion doublers, since the continuum
regularizations such as the dimensional regularization or the Pauli-Villars
regularization also produce a non-zero $\delta S$.
Therefore, the non-zero $\delta S$ reported in ref. 3 is {\it not}
a sign for the non-decoupling of fermion doublers.
So far there is no evidence against the decoupling of doublers in
the Wilson-Yukawa formulation.

We note that there is no reason to remove $\delta S^{1-loop}$ by a counter
term in the gauge-invariant regularizations we used above.
On the other hand, we {\it can} add a gauge invariant counter term of the form
$ tr [\tau^3 g^\dagger W_{\mu\nu}^a\tau^a g] B^{\mu\nu}$
to the Lagrangian, since the
dimension of this operator is 4. By changing the coefficient of this
term, we can obtain any value for the S parameter, including the
particular value
$S_{renormalized}^{1-loop} = \displaystyle {1\over 6\pi}$.
Thus this regularization scheme can not predict a definite
value for $S$ in perturbation theory,
and
we have to consider the problem non-perturbatively, for example, on the
lattice. If the term $ tr [\tau^3 g^\dagger W_{\mu\nu}^a\tau^a g]
B^{\mu\nu}$ is
relevant at the critical point where a continuum limit is taken,
we have to add this term to the Lagrangian and tune its coefficient, so that
the continuum limit can be defined at the critical point. There is no unique
prediction for the value of the S parameter in this case.
On the other hand, if
this term is irrelevant at the critical point
the value of the S parameter should be unique in the continuum limit once
all relevant parameters are fixed from experiments.

\vskip 0.5cm

Next we consider gauge non-invariant regularization schemes which
yield a manifestly renormalizable perturbative expansion.
The regulator part of the Lagrangian takes the form
$$\eqalign{
L_{reg}^{(dim)} & =  \bar\psi \sum_{\mu = 5}^D\gamma^\mu \
D_\mu \psi  \cr
L_{reg}^{(PV)} & = \sum_i \bar\psi_i (\gamma\cdot D + M_i +
yv\ g^R\cdot g^L )\psi_i  \cr
L_{reg}^{(lat)} &
= -{ar}\sum_\mu\ \bar\psi D^R_\mu D^R_\mu \psi  \cr }. \eqno(12)
$$
Note that the
$g$ field is absent in $L_{reg}$ except in the Yukawa coupling term of the
Pauli-Villars fields where it is necessary to cancel some divergences.
The terms without $g$ break gauge invariance explicitly.
Gauge anomalies as well as local
gauge non-invariant terms are generated, which have to
be removed
order by order in perturbative expansion by local counter
terms in order to
recover gauge( BRST) invariance.
This scheme gives a renormalizable perturbative expansion.

At the fermion 1-loop level, we can take the unitary gauge so that
the calculation of $S$ is identical to the gauge invariant case.
Therefore we obtain
$
\delta S^{1-loop} = - \displaystyle {1\over 6\pi}
$
for the dimensional and the Pauli-Villars regularization$^8$,
and
$
\delta S^{1-loop} = F(r)
$
for the lattice regularization$^3$.
However, these $\delta S^{1-loop}$ are not gauge invariant.
We have to remove $\delta S$ by a gauge non-invariant counter term
$ W_{\mu\nu}^3 B^{\mu\nu}$ in order to keep
gauge( BRST ) invariance of the renormalized theory. Then we obtain
$S_{renormalized}^{1-loop} =S_0^{1-loop} =\displaystyle {1\over 6\pi}$
independent of the choice of regulators.
This procedure also gives $ S^{1-loop}_{renormalized} = 0$
in the symmetric phase ($v=0$) and/or at $y=0$, see eq. (7).
However, it is not clear whether this way of calculation
in perturbation theory gives a gauge-fixing independent
and regularization independent result for the renormalized S parameter
at higher order of loops. It is necessary to prove the uniqueness
of the renormalized S parameter in the renormalizable perturbation theory.

\vskip 0.5cm

Finally we discuss other possible regularizations. So far we implicitly
required that the regularization preserves
vector gauge invariance such as SU(3)$_{color}$ or U(1)$_{EM}$.
We consider the following two types of regularizations which violate
this requirement. Since there is no physical Higgs field for SU(3)$_{color}$
and U(1)$_{EM}$, there is
no {\it gauge-invariant} scheme for these types of regularizations.

The first type uses Majorana terms in the regulator Lagrangian$^{9}$:
$$\eqalign{
L_{reg}^{(dim)} & = {1\over 2}\sum_{\mu =5}^D \left(
\psi^LC^L\gamma_\mu^L\partial^\mu\psi^L
+\bar\psi^L\gamma_\mu^RC^R\partial^\mu\bar\psi^L
+\psi^RC^R\gamma_\mu^R\partial^\mu\psi^R
+\bar\psi^R\gamma_\mu^LC^L\partial^\mu\bar\psi^R \right) \cr
L_{reg}^{(PV)} & = \sum_i \{ \bar\psi_i \left( \gamma\cdot D +
yv\ g^R\cdot g^L \right) \psi_i  \cr
& \qquad + {M_i\over 2}
(\psi_i^LC^L\psi_i^L-\bar\psi_i^LC^L\bar\psi_i^L-
\psi_i^RC^R\psi_i^R+\bar\psi_i^RC^R\bar\psi_i^R) \} \cr
L_{reg}^{(lat)} & = -{ar\over 2} \{
\psi^L C^L \triangle \psi^L
-\bar\psi^L C^L \triangle \bar\psi^L
-\psi^R C^R \triangle \psi^R
+\bar\psi^R C^R \triangle \bar\psi^R  \} \cr}. \eqno(13)
$$
Here the charge conjugation matrix $C$ satisfies
$ C \gamma_\mu C^{-1} = - \gamma_\mu^T$,
$C =C^LP_R + C_RP_L $,
$\gamma_\mu =\gamma_\mu^L P_L + \gamma_\mu^R P_R $, and
$\triangle = \sum_\mu \partial^\mu \partial_\mu$.
We find that all these regularizations give $\delta S^{1-loop} = 0$.
The fermion propagator in momentum space is given by
$$
<\psi^L \bar\psi^R>(p) = <\psi^R\bar\psi^L>(p) =
\cases{ \displaystyle -{m\over p^2_D + m^2} &
(14-a) \cr
\displaystyle -{m\over p^2 +m^2 + M_i^2} &
(14-b) \cr
\displaystyle -{m\over s^2(p)+m^2+M(p)^2} &
(14-c) \cr}
$$
for the dimensional regularization, the
i-th Pauli-Villars field, and the lattice regularization, respectively,
with $M(p) = \displaystyle {r\over a}\sum_\mu (1-\cos p_\mu )$,
$s^2(p) = a^{-2}\sum_\mu \sin^2(p_\mu a)$ and
$ p^2_D = \sum_{\mu =1}^D p_\mu^2$.
Due to the factor $m$ in the above propagators, the 1-loop contribution to
the S parameter
has a factor $m^2$, therefore $S^{1-loop} = 0$ at $ m = 0$.
This $m^2$ factor also makes the momentum integral for
$\Pi_{3Y}(p)$ more convergent,
so that $\Pi_{3Y}(0) = m^2 \times$  ( logarithmic divergent ) and
$ S^{1-loop} = m^2 \times$
( finite integral ).
Since the finite integral that appears for $S^{1-loop}$ should have a form of
$\displaystyle {1\over m^2} \times$ constant
by dimension analysis, only the {\it infrared} domain of
the integral contributes
to the S parameter, so that the S parameter in this case can not
depend on the form of an {\it ultra-violate} regularization.
We checked that this argument is true in an explicit calculation.
It is noted that fermion propagators
$<\psi^L \bar\psi^R>$ and $<\psi^R \bar\psi^L>$ in
the previous regularizations (5) and (12)
have extra terms without the explicit $m$ factor.
This is the reason why we obtained non-zero $\delta S^{1-loop}$
for these regularizations.

The second type of regularizations$^{10}$ introduces gauge singlet partners
$\chi^R$ for $\psi^L$ and $\chi^L$ for $\psi^R$;
$$\eqalign{
L_{reg}^{(dim)} & = \sum_{\mu =1}^D \bar\chi\gamma_\mu\partial^\mu\chi
+\sum_{\mu =5}^D [\bar\psi \gamma_\mu\partial^\mu\chi +
\bar\chi \gamma_\mu\partial^\mu\psi ]  \cr
L_{reg}^{(dim)} & = \sum_i \{ \bar\psi_i [\gamma\cdot D +
yv g^R\cdot g^L ]\psi_i
+ \bar\chi_i \gamma\cdot\partial \chi_i
+ M_i (\bar\chi_i\psi_i + \bar\psi_i\chi_i ) \}  \cr
L_{reg}^{(lat)} & = \bar\chi\gamma\cdot\partial \chi
-ar\sum_\mu \left(
\bar\psi \triangle \chi
+\bar\chi \triangle \psi \right)  \cr} . \eqno(15)
$$
The fermion propagator is given by
$$
<\psi^L \bar\psi^R>(p) = <\psi^R\bar\psi^L>(p) =
-m\times \cases{ \displaystyle {p^2_4\over (p^2_D)^2 + m^2p^2_4} &
(16-a) \cr
\displaystyle {p^2\over (p^2 + M_i^2)^2+p^2m^2} &
(16-b) \cr
\displaystyle { s^2(p)\over (s^2(p)+M(p)^2)^2+s^2(p)m^2} &
(16-c)
\cr} .
$$
Due to the extra factor $m$ in the propagator, these regularizations also
give $\delta S^{1-loop} = 0$. The explicit calculation confirmed
this.

Although the two types of regularizations (13) and (15) give
$\delta S^{1-loop} =0$, it is not so clear whether
$\delta S^{n-loops} = 0$ for all $n \geq 2$.
As pointed out before, it is necessary and important to prove
the uniqueness of the renormalized S parameter,
imposing gauge (BRST) invariance for the renormalized theory order by order
in the perturbative expansion.

\vskip 0.5cm
I would like to thank Prof. J. Shigemitsu, Prof. L. Randall and
Prof. A. Ukawa for useful discussion.
I would like to acknowledge the warm hospitality of Department of Physics,
University of Rome ''La Sapienza'', where this work was initiated.
\vskip 1cm
\noindent {\bf References}
\vskip 0.5cm

\item{1.}P.D. Swift, \pl{145}{1984}{256};
J. Smit, {\it Acta Phys. Polon.} {\bf B17}(1986)531;
\item{  }S. Aoki, 
\prl{60}{1988}{2109}; \pr{38}{1988}{618}; 
\item{  }K. Funakubo and T. Kashiwa, \prl{60}{1988}{2133}.
\item{2.}S. Aoki, I-H. Lee, and S.-S. Xue, {\it BNL Report} {\bf 42494}
(Feb. 1989); \pl{229}{1989}{79};
W. Bock, A.K. De, K. Jansen, J. Jersak, T. Neuhaus, and J. Smit,
\pl{232}{1989}{436}; \np{344}{1990}{207};
\item{  }S. Aoki, I-H. Lee, J. Shigemitsu, and R.E. Shrock,
\pl{243}{1990}{403};
\item{  }M. Golterman and D. Petcher, \pl{247}{1990}{370};
\item{  }S. Aoki, I-H. Lee, and R.E. Shrock,
\np{355}{1991}{383}.
\item{3.}M.J. Dugan and L. Randall,
preprint, MIT-CTP\#2050, HUTP-92/A001 (1992). 
\item{4.}M. Peskin and T. Takeuchi, \prl{65}{1990}{964};
B. Holdom and J. Terning, \pl{247}{1990}{88};
M. Golden and L. Randall, \np{361}{1991}{3}.
\item{5.}G. 't Hooft and M. Veltman, \np{44}{1972}{189}.
\item{6.}S. Aoki, I-H. Lee, and R.E. Shrock, \pr{45}{1992}{R13}.
\item{7.}S. Aoki, {\it Mod. Phys. Lett.} {\bf A5}(1990)2607;
\pr{42}{1990}{2806}.
\item{8.}S. Aoki, \pl{247}{1990}{357}.
\item{9.}T. Banks,
preprint, RU-91-13;
\item{   }S. Aoki, preprints, UTHEP-224(1991), UTHEP-230(1992).
\item{10.}A. Borrelli, L.Maiani, G. Rossi, R. Sisto and M. Testa,
\pl{221}{1989}{360}; \np{333}{1990}{335};
\item{   }Y. Kikukawa, preprint, DPNU-91-48, {\it Mod. Phys. Lett.} A in press.
\vfill
\end